\documentclass[a4paper]{article}

\usepackage{INTERSPEECH2019}
\usepackage{amsmath, graphicx}
\usepackage{hyperref, url}
\usepackage{amsfonts}
\usepackage{tikz}
\usepackage{subcaption}
\usepackage{float}
\usepackage{comment}
\usepackage{algpseudocode}
\usepackage[ruled]{algorithm}
\usepackage{pgfplots}

\usetikzlibrary{arrows, automata, positioning, shadows, calc, 3d}
\pgfdeclarelayer{background}
\pgfdeclarelayer{foreground}
\pgfsetlayers{background,main,foreground}
\graphicspath{ {imgs/} }
\tikzset{
    >=stealth',
    pre/.style={
           ->,
           ultra thick,
           shorten <=8pt,
           shorten >=8pt,},
    post/.style={
           <-,
           ultra thick,
           shorten <=1pt,
           shorten >=1pt,},
    bidirect/.style={
           <->,
           ultra thick,
           shorten <=1pt,
           shorten >=1pt,},
    ashadow/.style={
           opacity=.5,
           shadow xshift=0.07,
           shadow yshift=-0.07,},
    node style1/.style={
           fill=blue!20!white, drop shadow=ashadow,
           drop shadow=ashadow,},
    node style2/.style={
           fill=green!20!white, drop shadow=ashadow,
           drop shadow=ashadow,},
    node style3/.style={
           fill=red!20!white, drop shadow=ashadow,
           drop shadow=ashadow,},
}

\title{Generative Adversarial Networks for Unpaired Voice Transformation \\ on Impaired Speech}
\name{Li-Wei Chen$^1$, Hung-Yi Lee$^2$, Yu Tsao$^3$}

\address{
  $^1$$^2$National Taiwan University,   $^3$Academia Sinica}
\email{b04901014@ntu.edu.tw, hungyilee@ntu.edu.tw, yu.tsao@citi.sinica.edu.tw}

\begin{document}
\maketitle

\begin{abstract}
This paper\footnote{This work was supported in part by Ministry of Science and Technology (MOST), R.O.C. and NVIDIA.} focuses on using voice conversion (VC) to improve the speech intelligibility of surgical patients who have had parts of their articulators removed.
Due to the difficulty of data collection, VC without parallel data is highly desired. 
Although techniques for unparallel VC---for example, CycleGAN---have been developed, they usually focus on transforming the speaker identity, and directly transforming the speech of one speaker to that of another speaker and as such do not address the task here.
In this paper, we propose a new approach for unparallel VC.
The proposed approach transforms impaired speech to normal speech while preserving the linguistic content and speaker characteristics.
To our knowledge, this is the first end-to-end GAN-based unsupervised VC model applied to impaired speech. 
The experimental results show that the proposed approach outperforms CycleGAN.

\end{abstract}


\section{Introduction}
\label{sec:intro}

Voice conversion (VC) is a task aimed at converting the speech signals from a certain acoustic domain to another while keeping the linguistic content the same.  
Examples of acoustic domains include not only speaker identity~\cite{stylianou1998continuous, kain1998spectral, saito2011one, kinnunen2017non}, but many other factors orthogonal to the linguistic content, such as speaking style, speaking rate~\cite{rentzos2003transformation}, noise condition, emotion~\cite{aihara2012gmm, kawanami2003gmm}, and accent~\cite{oyamada2017non}, with potential applications ranging from speech enhancement~\cite{kain2007improving, rudzicz2011acoustic}, computer-assisted pronunciation training for non-native language learner~\cite{oyamada2017non}, speaking assistance~\cite{nakamura2012speaking}, to name a few.

This paper focuses on using VC to improve the speech intelligibility of surgical patients who have had parts of their articulators removed.
Because of the removal of parts of the articulator, a patient's speech may become distorted and difﬁcult to understand. 
VC methods can be applied to convert the distorted speech such that it is clear and more intelligible.
In this work, we consider a VC model without ASR~\cite{zerospeech} because collecting a large amount of data to train an ASR for impaired speech is laborious and not practical.

Non-negative matrix factorization (NMF) based VC has been used for this task~\cite{NMFVC,NMF1,NMF2}.  
In previous work, paired utterances from both patients and unimpaired people were needed for training. 
Collecting a large amount of audio from patients is difficult under this task because even speaking for a long time is usually difficult for them, not to mention the collection of paired data.
Due to the lack of training data, to our best knowledge, deep learning has not been widely applied on this task yet.

\begin{figure}[thb]
   \centering
   \includegraphics[width=\linewidth]{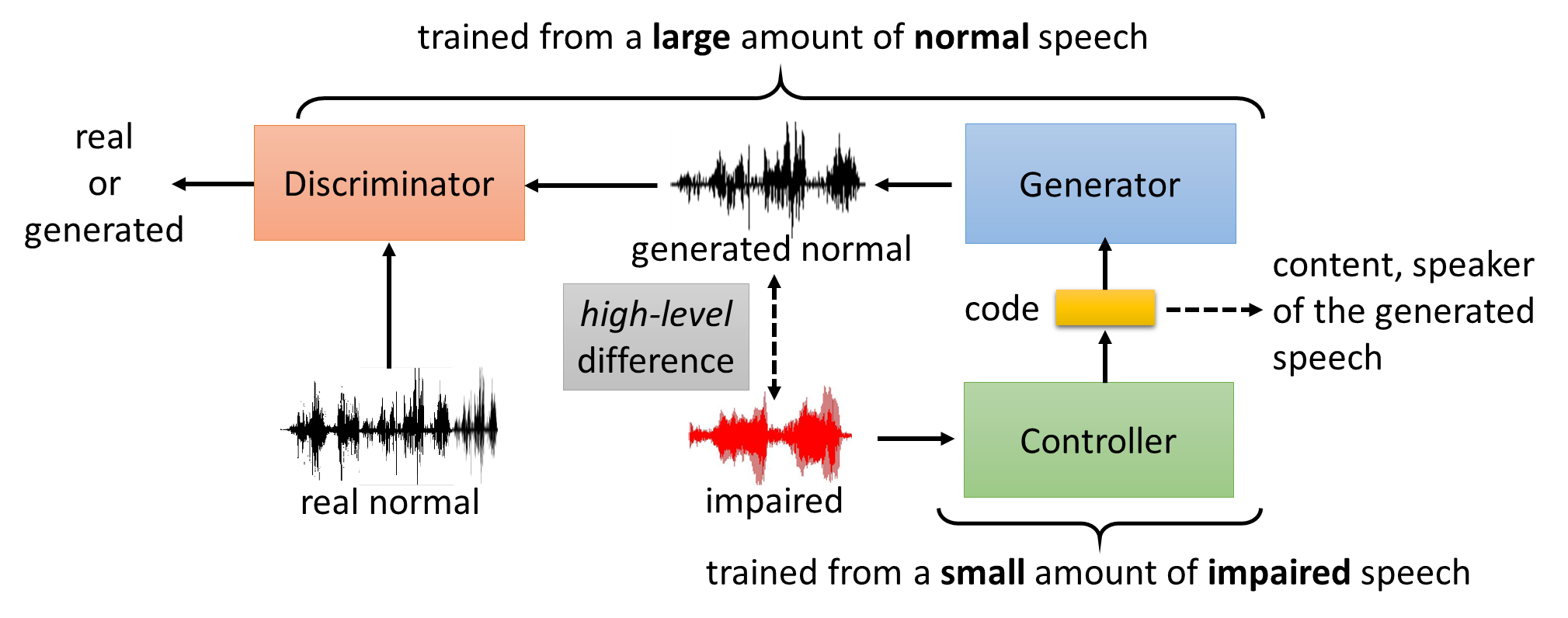}
   \caption{Overview of proposed approach. The difference is evaluated by a network instead of using low-level signal differences.
   The controller learns to minimize the high-level difference.} 
   \label{fig:overview}
\end{figure}

After the success of deep learning in various domains, many researchers have attempted to incorporate deep learning into the VC framework, but most focus on speaker identity conversion.  
Most previous work requires aligned data, but due to the difficulties in obtaining aligned data, approaches utilizing generative models such as variational autoencoders (VAEs)~\cite{VAE1,VAE2} and generative adversarial networks (GANs)~\cite{GAN1,GAN2} were studied because they can be trained with non-parallel data.
VC for articulation disorders without parallel data is highly desired due to the difficulty of data collection. 
To achieve that, one can simply apply the techniques developed for speaker identity VC by considering the patient with the articulation disorder as the source speaker, and the unimpaired person as the target speaker.
However, the model thus learned would simply convert the voice of the source speaker into that of the target speakers without preserving the source speaker's individuality.
Even worse, the speaker VC model may change only speaker characteristics, but yield a converted voice that is still unclear. 
Therefore, to achieve VC for articulation disorders without parallel data, a new approach must be developed.\\\indent
The overview of the proposed approach is shown in Figure~\ref{fig:overview}. 
The proposed model includes a generator, a discriminator, and a controller. 
The generator and discriminator form a GAN which is learned from a large amount of normal speech which is easier to collect than impaired speech. 
The discriminator learns to judge whether the input is real speech or if it has been generated by the generator.
The generator takes a code which represents the content and speaker of the audio to be generated as input, and generates normal speech to fool the discriminator.
The impaired speech is used only to train the controller. 
Given impaired speech as input, the controller outputs a code which is taken as the input of the generator, and the generator generates normal speech based on the input code.
The controller learns to generate code that makes the generator output normal speech with the same linguistic content and the same speaker characteristics as the impaired speech, thus minimizing their high-level differences. 

To guide the controller learning, we require automatic ways to evaluate this high-level difference.
Inspired by perception loss, widely used in image processing~\cite{Percept}, we use the hidden layers of the discriminator to evaluate the similarity of two audio segments.
Compared with CycleGAN, which maps from the source speaker to the target speaker also in an unsupervised way, the proposed approach better improves speech intelligibility while preserving speaker characteristics. 

\section{Proposed Approaches}
\label{sec:uvt}

The proposed approach consists of three models: a generator $G$, a discriminator $D$, and a controller $C$.
For training data, we have a large amount of speech from unimpaired subjects: $\mathcal{T} = \{x_i^t\}_{i=1}^N$, where $x^t_i$ is a fixed-length acoustic feature sequence from the utterances of unimpaired subjects, and $N$ is the number of audio segments in the training set.
We also have the speech of a patient, $\mathcal{S} =  \{x_i^s\}_{i=1}^{N^\prime}$, where $N^\prime$ is the number of audio segments for the patient, but the data size is much smaller than that of unimpaired subjects ($N^\prime \ll N$).

The content of normal speech and impaired speech are completely unrelated.
During testing, given an utterance of the patient, it is first equally segmented into a sequence of audio segments.
The controller takes the audio segments as input, and the generator transforms them into normal speech.

\subsection{Generator-Discriminator}
\label{ssec:GAN}

We use the audio of unimpaired subjects $\mathcal{T}$ to train the audio generator $G$ and discriminator $D$.
The generator is used to generate audio $\tilde{x}$ given a vector $c$, that is, $\tilde{x} = G(c)$, and the discriminator $D$ attempts to distinguish $x^t \sim \mathcal{T}$ from  $\tilde{x} \sim G$ while the generator tries to fool it. 
As shown below, the objective functions for $D$ and $G$ follow the idea of LSGAN~\cite{LSGAN}:
\begin{align}
  \mathcal{L}_{D} & = \mathbb{E}_{x^t \sim \mathcal{T}}[ (D(x^t) - 1)^2 ] \label{eq:mylsgand} \\
                  & + \mathbb{E}_{c \sim P_c(c), \tilde{x} \sim G(c)}[ (D(\tilde{x}))^2 ] \nonumber \\
  \mathcal{L}_{G} & = \mathbb{E}_{c \sim P_c(c), \tilde{x} \sim G(c)}[ (D(\tilde{x}) - 1)^2 ] \label{eq:mylsgang}
\end{align}

In (\ref{eq:mylsgand}), $D$ learns to assign normal speech $x^t$ a score of one, and assign a score of zero to generated audio segments $\tilde{x}$.
At the same time, in (\ref{eq:mylsgang}), $G$ learns to generate an $\tilde{x}$ that yields a score of one from $D$.
Here $c$ is the output of the controller $C$, which we assume has a distribution $P_c(c)$ in (\ref{eq:mylsgand}) and (\ref{eq:mylsgang})\footnote{To be specific, $c \sim P_c(c)$ is equivalent to $c \sim C(x^s), x^s \sim \mathcal{S}$. This is made clear below.}. 
After the above training procedure, we have a generator $G$ which generates normal speech given a condition vector $c$.
The vector $c$ controls the generated audio of $G$.
By choosing the condition vector $c$ properly, we generate audio segments with the desired content and speaker characteristic. 
The core idea   is that  a large amount of normal speech can be used to train a generator $G$ which can generate  high quality speech, and the impaired speech is only used to select $c$, which is a much easier task than speech generation.

\def\cscale{0.58}
\def\nodedist{1.5cm}
\begin{figure*}
\centering
\begin{tikzpicture}[node distance=\nodedist, auto, scale=\cscale]

\newcommand{\flayer}[9]{
  \def \h{#2} 
  \def \w{#3} 
  \def \scaleh{#4} 
  \def \scalew{#5} 
  \def \scaled{#7} 
  \def \d{#9} 
  \def \factor{50} 
  \def \colorfill{#6}
  \def \labelpos{#8}

  \coordinate (A) at ($ #1 + (\d / \factor / \scaled, \h / \factor / \scaleh, \w / \factor / \scalew) $);
  \coordinate (B) at ($ #1 + (\d / \factor / \scaled, -\h / \factor / \scaleh, \w / \factor / \scalew) $);
  \coordinate (C) at ($ #1 + (\d / \factor / \scaled, -\h / \factor / \scaleh, -\w / \factor / \scalew) $);
  \coordinate (D) at ($ #1 + (\d / \factor / \scaled, \h / \factor / \scaleh, -\w / \factor / \scalew) $);
  \coordinate (E) at ($ #1 + (-\d / \factor / \scaled, \h / \factor / \scaleh, \w / \factor / \scalew) $);
  \coordinate (F) at ($ #1 + (-\d / \factor / \scaled, -\h / \factor / \scaleh, \w / \factor / \scalew) $);
  \coordinate (G) at ($ #1 + (-\d / \factor / \scaled, -\h / \factor / \scaleh, -\w / \factor / \scalew) $);
  \coordinate (H) at ($ #1 + (-\d / \factor / \scaled, \h / \factor / \scaleh, -\w / \factor / \scalew) $);
  
  \fill[\colorfill!50,opacity=0.3] (A) -- (B) -- (C) -- (D) -- cycle;
  \fill[\colorfill!50,opacity=0.5] (E) -- (F) -- (G) -- (H) -- cycle;
  \fill[\colorfill!50,opacity=0.3] (E) -- (A) -- (D) -- (H) -- cycle;
  \fill[\colorfill!50,opacity=0.3] (E) -- (A) -- (B) -- (F) -- cycle;
  \fill[\colorfill!50,opacity=0.5] (G) -- (C) -- (B) -- (F) -- cycle;
  \fill[\colorfill!50,opacity=0.5] (G) -- (C) -- (D) -- (H) -- cycle;
  
  \draw[\colorfill!70] (A) -- (B) -- (C) -- (D) -- cycle;
  \draw[\colorfill!70] (E) -- (F) -- (G) -- (H) -- cycle;
  \draw[\colorfill!70] (E) -- (A) -- (D) -- (H) -- cycle;
  \draw[\colorfill!70] (E) -- (A) -- (B) -- (F) -- cycle;
  \draw[\colorfill!70] (G) -- (C) -- (B) -- (F) -- cycle;
  \draw[\colorfill!70] (G) -- (C) -- (D) -- (H) -- cycle;
  \draw[\colorfill!70] (B) -- (C) node[below right, pos=\labelpos]{\scriptsize{\w}};
  \draw[\colorfill!70] (B) -- (F) node[below, pos=0.5]{\scriptsize{\d}};
  \draw[\colorfill!70] (F) -- (E) node[left, pos=\labelpos]{\scriptsize{\h}};
}

\newcommand{\blayer}[9]{
  \def \h{#2} 
  \def \w{#3} 
  \def \scaleh{#4} 
  \def \scalew{#5} 
  \def \scaled{#7} 
  \def \d{#9} 
  \def \factor{50} 
  \def \colorfill{#6}
  \def \labelpos{#8}

  \coordinate (A) at ($ #1 + (\d / \factor / \scaled, \h / \factor / \scaleh, \w / \factor / \scalew) $);
  \coordinate (B) at ($ #1 + (\d / \factor / \scaled, -\h / \factor / \scaleh, \w / \factor / \scalew) $);
  \coordinate (C) at ($ #1 + (\d / \factor / \scaled, -\h / \factor / \scaleh, -\w / \factor / \scalew) $);
  \coordinate (D) at ($ #1 + (\d / \factor / \scaled, \h / \factor / \scaleh, -\w / \factor / \scalew) $);
  \coordinate (E) at ($ #1 + (-\d / \factor / \scaled, \h / \factor / \scaleh, \w / \factor / \scalew) $);
  \coordinate (F) at ($ #1 + (-\d / \factor / \scaled, -\h / \factor / \scaleh, \w / \factor / \scalew) $);
  \coordinate (G) at ($ #1 + (-\d / \factor / \scaled, -\h / \factor / \scaleh, -\w / \factor / \scalew) $);
  \coordinate (H) at ($ #1 + (-\d / \factor / \scaled, \h / \factor / \scaleh, -\w / \factor / \scalew) $);
  
  \fill[\colorfill!50,opacity=0.3] (A) -- (B) -- (C) -- (D) -- cycle;
  \fill[\colorfill!50,opacity=0.5] (E) -- (F) -- (G) -- (H) -- cycle;
  \fill[\colorfill!50,opacity=0.3] (E) -- (A) -- (D) -- (H) -- cycle;
  \fill[\colorfill!50,opacity=0.3] (E) -- (A) -- (B) -- (F) -- cycle;
  \fill[\colorfill!50,opacity=0.5] (G) -- (C) -- (B) -- (F) -- cycle;
  \fill[\colorfill!50,opacity=0.5] (G) -- (C) -- (D) -- (H) -- cycle;
  
  \draw[\colorfill!70] (A) -- (B) -- (C) -- (D) -- cycle;
  \draw[\colorfill!70] (E) -- (F) -- (G) -- (H) -- cycle;
  \draw[\colorfill!70] (E) -- (A) -- (D) -- (H) -- cycle;
  \draw[\colorfill!70] (E) -- (A) -- (B) -- (F) -- cycle;
  \draw[\colorfill!70] (G) -- (C) -- (B) -- (F) -- cycle;
  \draw[\colorfill!70] (G) -- (C) -- (D) -- (H) -- cycle;
  \draw[\colorfill!70] (H) -- (E) node[left, pos=\labelpos]{\scriptsize{\w}};
  \draw[\colorfill!70] (D) -- (H) node[above, pos=0.5]{\scriptsize{\d}};
  \draw[\colorfill!70] (D) -- (C) node[right, pos=\labelpos]{\scriptsize{\h}};
}

\def \dist{3.2}
\def \hdist{4.5*1.3}

\blayer{(0, 0, 0)}{64}{128}{1.5}{1.5}{blue!60!black}{1}{0.2}{1};
\blayer{(\dist, 0, 0)}{32}{32}{0.8}{0.8}{blue!60!black}{4}{0.2}{64};
\blayer{(\dist * 2, 0, 0)}{16}{16}{0.6}{0.6}{blue!60!black}{7}{0.2}{128};
\blayer{(\dist * 3, 0, 0)}{8}{8}{0.4}{0.4}{blue!60!black}{10}{0.2}{256};
\blayer{(\dist * 4, 0, 0)}{4}{4}{0.3}{0.3}{blue!60!black}{13}{0.2}{512};
\blayer{(\dist * 5, 0, 0)}{2}{2}{0.2}{0.2}{blue!60!black}{18}{0.2}{1024};

\flayer{(0, 0, \hdist)}{64}{128}{1.5}{1.5}{green!60!black}{1}{0.1}{1};
\flayer{(\dist, 0, \hdist)}{32}{32}{0.8}{0.8}{green!60!black}{4}{0.1}{64};
\flayer{(\dist * 2, 0, \hdist)}{16}{16}{0.6}{0.6}{green!60!black}{7}{0.1}{128};
\flayer{(\dist * 3, 0, \hdist)}{8}{8}{0.4}{0.4}{green!60!black}{10}{0.1}{256};
\flayer{(\dist * 4, 0, \hdist)}{4}{4}{0.3}{0.3}{green!60!black}{13}{0.1}{512};
\flayer{(\dist * 5, 0, \hdist)}{2}{2}{0.2}{0.2}{green!60!black}{18}{0.1}{1024};

\blayer{(\dist * 5.1, 0, -\hdist)}{64}{128}{1.5}{1.5}{orange!60!red}{1}{0.1}{1};
\blayer{(\dist * 4, 0, -\hdist)}{64}{128}{1.5}{1.5}{orange!60!red}{1}{0.1}{5};
\blayer{(\dist * 3, 0, -\hdist)}{32}{32}{0.8}{0.8}{orange!60!red}{4}{0.1}{64};
\blayer{(\dist * 2, 0, -\hdist)}{16}{8}{0.6}{0.6}{orange!60!red}{7}{0.1}{128};
\blayer{(\dist * 1, 0, -\hdist)}{8}{4}{0.4}{0.4}{orange!60!red}{10}{0.1}{256};
\blayer{(\dist * 0, 0, -\hdist)}{4}{2}{0.3}{0.3}{orange!60!red}{13}{0.1}{512};

\draw[blue!70!red, pre, opacity=0.7] (0, 0, \hdist) -- (\dist - 64 / 50 / 4 / 2, 0, \hdist) node[below, pos=0.5]{};
\draw[blue!70!red, pre, opacity=0.7] (\dist, 0, \hdist) -- (\dist * 2 - 128 / 50 / 7 / 2, 0, \hdist) node[below, pos=0.5]{};
\draw[blue!70!red, pre, opacity=0.7] (\dist * 2, 0, \hdist) -- (\dist * 3 - 256 / 50 / 10 / 2, 0, \hdist) node[below, pos=0.5]{};
\draw[blue!70!red, pre, opacity=0.7] (\dist * 3, 0, \hdist) -- (\dist * 4 - 512 / 50 / 13 / 2, 0, \hdist) node[below, pos=0.5]{};
\draw[blue!70!red, pre, opacity=0.7] (\dist * 4 + 512 / 50 / 18 / 1.5, 0, \hdist) -- (\dist * 5 - 1024 / 50 / 18 / 1.5, 0, \hdist) node[below, pos=0.5]{};

\draw[black, pre, opacity=0.9] (\dist * 5.1  - 64 / 50 / 4 / 2, 0, -\hdist) -- (\dist * 4, 0, -\hdist) node[below, pos=0.54]{\shortstack{\scriptsize{\textbf{Add delta}} \\\scriptsize{\textbf{and }}\scriptsize{\textbf{delta-delta}}}};
\draw[blue!70!red, pre, opacity=0.7] (\dist * 4  - 64 / 50 / 4 / 2, 0, -\hdist) -- (\dist * 3, 0, -\hdist) node[below, pos=0.5]{};
\draw[blue!70!red, pre, opacity=0.7] (\dist * 3 - 128 / 50 / 7 / 2, 0, -\hdist) -- (\dist * 2, 0, -\hdist) node[above, pos=0.5]{};
\draw[blue!70!red, pre, opacity=0.7] (\dist * 2 - 128 / 50 / 10 / 2, 0, -\hdist) -- (\dist * 1, 0, -\hdist) node[above, pos=0.5]{};
\draw[blue!70!red, pre, opacity=0.7] (\dist * 1 - 256 / 50 / 13 / 2, 0, -\hdist) -- (\dist * 0 + 512 / 50 / 13 / 2, 0, -\hdist) node[above, pos=0.5]{};
\draw[green!50!purple, pre, opacity=0.7] (\dist * 0 - 512 / 50 / 18 / 1.5, 0, -\hdist) -- (\dist * -0.8, 0, -\hdist) node[above, pos=0.5]{};

\draw[gray!50!cyan, opacity=0.5] (\dist * 4  - 64 / 50 / 4 / 2, 0, -\hdist) edge[out=150, in=30, pre] node[above, pos=0.5]{} (\dist * 2 - 128 / 50 / 10 / 2, 0, -\hdist) ;
\draw[gray!50!cyan, opacity=0.5] (\dist * 4  - 64 / 50 / 4 / 2, 0, -\hdist) edge[out=150, in=30, pre] node[above, pos=0.5]{} (\dist * 1 - 256 / 50 / 13 / 2, 0, -\hdist) ;
\draw[gray!50!cyan, opacity=0.5] (\dist * 4  - 64 / 50 / 4 / 2, 0, -\hdist) edge[out=150, in=30, pre] node[above, pos=0.5]{} (\dist * 0 + 512 / 50 / 13 / 2, 0, -\hdist) ;

\draw[red!50!black, pre, opacity=0.7] (\dist * 4 - 512 / 50 / 13 / 2, 0, 0, 0) -- (\dist * 3, 0, 0) node[above, pos=0.5]{};
\draw[red!50!black, pre, opacity=0.7] (\dist * 3 - 256 / 50 / 10 / 2, 0, 0) -- (\dist * 2, 0, 0) node[above, pos=0.5]{};
\draw[red!50!black, pre, opacity=0.7] (\dist * 2 - 128 / 50 / 7 / 2, 0, 0) -- (\dist * 1, 0, 0) node[above, pos=0.5]{};
\draw[red!50!black, pre, opacity=0.7] (\dist * 1 - 64 / 50 / 4 / 2, 0, 0) -- (0, 0, 0) node[above, pos=0.5]{};
\draw[red!50!black, pre, opacity=0.7] (\dist * 5 - 1024 / 50 / 18 / 1.5, 0, 0) -- (\dist * 4 + 512 / 50 / 18 / 1.5, 0, 0) node[below, pos=0.5]{};

\draw[black, opacity=0.9] (\dist * 5, 0, \hdist) edge[out=10, in=-70, pre] node[below right, pos=0.5]{\shortstack{\scriptsize{\textbf{Projection onto}}\\\scriptsize{\textbf{unit ball}}}} (\dist * 5, 0, 0);

\node (styleA) at (-\dist / 2, 0, \hdist) {\shortstack{\textbf{\scriptsize{Impaired}}\\\textbf{\scriptsize{speech}}}};
\node (styleB) at (-\dist / 2, 0, 0) {\shortstack{\textbf{\scriptsize{Normal}}\\\textbf{\scriptsize{speech}}}};
\node (styleB) at (-\dist / 1.2, 0, -\hdist) {\shortstack{\textbf{\scriptsize{Scalar}}\\\textbf{\scriptsize{score}}}};
\node (styleB) at (\dist * 5, 0, \hdist - 1.5) {\textbf{c}};

\matrix[draw, thick, opacity=0.8](m1) at ($ (current bounding box.east) + (0.8, -2.5) $) {
  \node[shape=rectangle, fill=blue!70!red, opacity=0.7, line width=1, label=right:\scriptsize{Conv2D}] {}; \\
  \node[shape=rectangle, fill=red!50!black, opacity=0.7, line width=1, label=right:\scriptsize{Deonv2D}] {}; \\
  \node[shape=rectangle, fill=gray!50!cyan, opacity=0.7, line width=1, label=right:\scriptsize{Maxpooling}] {}; \\
  \node[shape=rectangle, fill=green!50!purple, opacity=0.7, line width=1, label=right:\scriptsize{Dense Layer}] {}; \\
};
\node at ($(m1.north) + (0, 0.5)$) {\textbf{Color for Arrows}};

\matrix[draw, thick, opacity=0.8](m2) at ($ (current bounding box.west) + (1.0, 2.5) $) {
  \node[shape=rectangle, fill=green!60!black, opacity=0.7, line width=1, label=right:\scriptsize{Controller}] {}; \\
  \node[shape=rectangle, fill=blue!60!black, opacity=0.7, line width=1, label=right:\scriptsize{Generator}] {}; \\
  \node[shape=rectangle, fill=orange!60!red, opacity=0.7, line width=1, label=right:\scriptsize{Discriminator}] {}; \\
};
\node at ($(m2.north) + (0, 0.5)$) {\textbf{Color for Blocks}};

\node[canvas is zy plane at x=0, opacity=0.9] (img) at (0, 0, \hdist) {\includegraphics[scale=0.49]{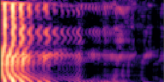}};
\node[canvas is zy plane at x=0, opacity=0.9] (img) at (0, 0, 0) {\includegraphics[scale=0.49]{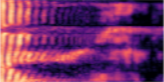}};
\node[canvas is zy plane at x=\dist*2.42*0.968, opacity=0.4] (img) at (0, 0, -\hdist) {\includegraphics[scale=0.49]{mel1}};
\node[canvas is zy plane at x=\dist*2.41*0.968, opacity=0.4] (img) at (0, 0, -\hdist) {\includegraphics[scale=0.49]{mel1}};
\node[canvas is zy plane at x=\dist*2.40*0.968, opacity=0.4] (img) at (0, 0, -\hdist) {\includegraphics[scale=0.49]{mel1}};
\node[canvas is zy plane at x=\dist*2.39*0.968, opacity=0.4] (img) at (0, 0, -\hdist) {\includegraphics[scale=0.49]{mel1}};
\node[canvas is zy plane at x=\dist*2.38*0.968, opacity=0.4] (img) at (0, 0, -\hdist) {\includegraphics[scale=0.49]{mel1}};
\node[canvas is zy plane at x=\dist*3.06*0.968, opacity=0.9] (img) at (0, 0, -\hdist) {\includegraphics[scale=0.49]{mel1}};
\end{tikzpicture}
\caption{Controller-generator-discriminator configuration}
\label{fig:Model}
\end{figure*}

\subsection{Controller}
\label{ssec:CGD}

Given the audio segment $x^s$ from a patient, we want to find its corresponding counterpart $x^t$ in the domain of normal speech.
The basic idea is to properly choose the condition $c$ that causes $G$ to generate speech $x^t$ similar to $x^s$.
If $x^t$ is close to $x^s$, they may contain the same linguistic information with the same speaker characteristics, but the $x^t$ generated by $G$ sounds like normal speech (which is what we want) because $G$ has learned to generate normal speech.

The controller $C$ takes an audio segment $x^s$ as input, and outputs its corresponding condition $c$ as the input of $G$.
Here we assume only a small amount of audio $\mathcal{S}$ from the patient is available as training data.
$\mathcal{S}$ is used only to train controller $C$.
$C$ is learned by minimizing the following loss:
\begin{equation}
\mathcal{L}_{C} = \mathbb{E}_{x^s \sim \mathcal{S}} [ L( G(C(x^s)), x^s ) ]   
\label{eq:att}
\end{equation}
The metric $L(.,.)$ is used to evaluate the difference between two audio segments.
In (\ref{eq:att}), $C$ learns to make the input $x^s$ and the corresponding output of the generator $G(C(x^s))$ as close as possible.
$L(.,.)$ is defined in the next subsection. 
If we jointly optimize $G$ and $C$, minimizing (\ref{eq:att}) is equivalent to training an auto-encoder (the controller is an encoder, while the generator is a decoder).
However, we only update $C$ when we minimize (\ref{eq:att}). 
This is very critical for the success of this approach, 
because if $G$ is also updated to minimize (\ref{eq:att}), we cannot guarantee that $G$ still generates normal speech after the update.

\subsection{Distance Measure for Audio}

For distance $L(.,.)$, both L1 and L2 loss are not suitable because we seek to evaluate the similarity of the content and the speaker characteristics between two audio segments, not merely low-level signal similarity.
On image style transfer tasks, the perceptual loss \cite{Percept}, which utilizes the layers of a CNN classifier as features and applies a distance measure to these, has been shown to produce finer results than pixel-wise loss.
Here we borrow this idea to evaluate high-level audio similarity. 
Instead of training another classifier, we use the discriminator as the objective classifier for the distance measure. 

For the perceptual loss, we choose the Laplacian pyramid Lap$_1$ loss~\cite{GLO}. 
We use the notation $D_l(x)$ to denote the output of the \mbox{$l$-th} layer of the discriminator $D$ given input $x$. 
Then $L(.,.)$ in (\ref{eq:att}) is formulated as
\begin{align}
  L(x,x^\prime) & = \sum_l 2^{-2l} \lvert D_l(x) - D_l(x^\prime)\rvert_1 \label{eq:mygand1}
\end{align}
The L1 distance of the hidden layer output $D_l(x)$ is computed.
In (\ref{eq:mygand1}), all the hidden layers of $D$ are considered to capture information at different granularities.
The weights for each layer follows \cite{GLO}.

\section{Implementation}

\subsection{Acoustic Feature Processing}
\label{ssec:feat}
We use the mel spectrogram as the input of the controller and the output of the generator. 
Before transforming into the spectrogram, we trim audio silence and perform volume normalization.
All audio is converted to a 16kHz sample rate. 
After that, we use a 50 milliseconds window length, a 12.5 milliseconds hop length, and a 1024 FFT window size for the STFT. 

After constructing the spectrogram, we construct the mel spectrogram using 128 mel frequency bands with a frequency range from 55Hz to 7600Hz.
Then we turn the mel spectrogram into the decibel scale and standardize the features across the time dimension to zero mean and unit variance. 
We clip the values between $-c$ and $c$.
Although the hyperparameter $c$ is somewhat data dependent, we find that $c = 3$ works well in most cases. 
Using these settings, most of the human speech mel spectrograms can be transformed back to the raw waveforms with little audible distortion. 
Since the feature values are constrained to $[-c, c]$, we use $c \cdot tanh(\cdot)$ as the output activation for our generator.

To convert the mel spectrogram back to the raw waveform, we first rescale the model output and then multiply the psuedo inverse of the mel filter bank to recover the linear spectrogram. 
Finally, using Griffin and Lim's algorithm to estimate the phase, we reconstruct the raw waveform.

\subsection{Model Configuration}
\label{ssec:modconfig}

The network structure of the discriminator, generator, and controller is shown in Figure~\ref{fig:Model}.
So that the discriminator better captures dynamics, we augment the input spectrogram with the first and second order deltas in both time and frequency dimensions.
The time-frequency bin of the original spectrum is represented by a scalar; after augmentation, each bin is a 5-dimensional vector.
To make gradient propagation more efficient, we augment each hidden layer of the discriminator with its input. However, since the discriminator uses strided convolutions, the spacial dimension of the input and hidden layers will be inconsistent.
To overcomes this, we apply max pooling on the input to make the spacial dimension consistent with hidden layers, then augment them as additional channels to the hidden layers.

The pool size and stride are determined by the shape of the corresponding discriminator layer.
Before $c$ is fed into the generator, we project $c$ onto the $L_2$ unit ball as:
\begin{equation}
  c \leftarrow \frac{c}{max(\lVert c\rVert_2, 1)} \label{eq:myc}  
\end{equation}
In the real-world implementation, $\mathcal{L}_D$, $\mathcal{L}_G$, and  $\mathcal{L}_C$ are updated once in each training iteration.

\subsection{Training Details}
\label{ssec:tra}
We trained the proposed model using the Adam optimizer with a 0.0002 learning rate, and $\beta_1=0.5$ and $\beta_2=0.9$ for the controller and generator.
A learning rate of 0.0001 was applied to the discriminator to avoid fast convergence.
Instance normalization~\cite{InsN} and spectral normalization~\cite{SN} were applied to both the generator and discriminator to stabilize training.
ELU activation~\cite{ELU} was used for the generator and controller, and RELU activation was used for the discriminator.
We used a batch size of 64, a dropout rate of 0.9 for the controller, and 0.8 for the discriminator.
Further details may be found in our implementation code: \url{https://github.com/b04901014/ISGAN}.

\section{Experiments}
\label{sec:exp}

\subsection{Experimental Setup}

The normal utterances for training the generator and discriminator came from the ASTMIC dataset, which contains 100 male speakers with different speaking rates, each reading out 200 different Mandarin sentences.
The impaired speech, from an orally-impaired male speaker, included 132 utterances in Mandarin.
As this speaker had a serious articulatory injury, most of his utterances were unintelligible to normal people.

We compare the proposed approach with two baselines:
\begin{itemize}
    \item Conditional GAN (cGAN): As this method is supervised, parallel audio is needed. A normal male speaker reads the same sentences as the patient, and the paired data is obtained by aligning the utterances of the patient and the normal speaker. cGAN involves training a discriminator and a conditional generator which takes impaired speech as input and outputs normal speech. Here the cGAN training algorithm is that from Pix2pix~\cite{Pix2pix}. The network architecture of the discriminator in cGAN is the same as in the proposed approach, and the network architecture of the conditional generator is the same as the cascaded controller and generator in the proposed approach. 
    \item CycleGAN~\cite{CycleGAN}: As with the proposed approach, this method needs no parallel data.
    With data from two domains, CycleGAN learns a generator to transform an object from one domain to another domain.
    Here we consider the ASTMIC audio as one domain and the audio of the patient as the other domain. 
    The discriminator used in CycleGAN also has the same network structure as the proposed approach.
    Likewise, the generator in CycleGAN is the concatenation of the controller and the generator in the proposed approach.
\end{itemize}

\begin{figure*}
\centering
\begin{subfigure}[t]{.18\textwidth}
  \vskip 0pt
  \centering
  \includegraphics[scale=0.25]{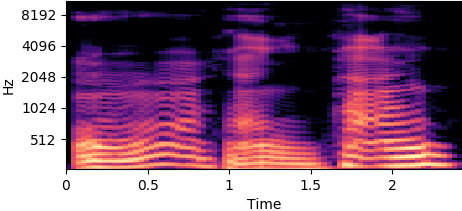}
  \caption{Impaired speech}
  \label{fig:SpecOR}
\end{subfigure}
\begin{subfigure}[t]{.18\textwidth}
  \vskip 0pt
  \centering
  \includegraphics[scale=0.25]{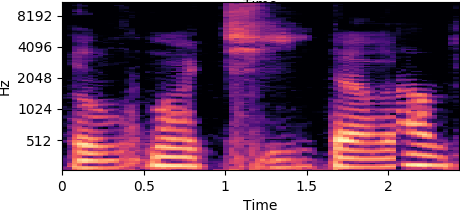}
  \caption{Proposed model}
  \label{fig:full}
\end{subfigure}
\begin{subfigure}[t]{.18\textwidth}
  \vskip 0pt
  \centering
  \includegraphics[scale=0.25]{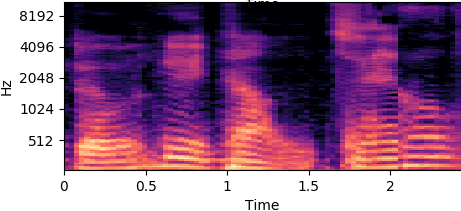}
  \caption{NoSD model}
  \label{fig:SpecNSD}
\end{subfigure}
\begin{subfigure}[t]{.18\textwidth}
  \vskip 0pt
  \centering
  \includegraphics[scale=0.25]{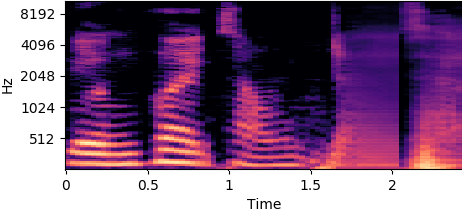}
  \captionsetup{justification=centering}
  \caption{CycleGAN\newline transformed}
  \label{fig:Cyctrans}
\end{subfigure}
\begin{subfigure}[t]{.18\textwidth}
  \vskip 0pt
  \centering
  \includegraphics[scale=0.25]{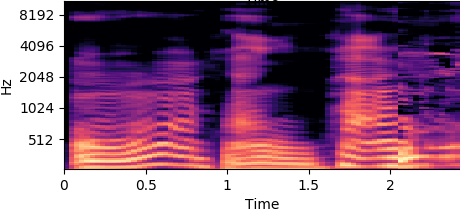}
  \captionsetup{justification=centering}
  \caption{CycleGAN\newline reconstructed}
  \label{fig:Cycrec}
\end{subfigure}
\caption{Spectrogram of impaired speech before and after transformation by each model. (a) impaired speech, (b) transformed speech using proposed method, (c) discriminator without skip connection (NoSD), (d) transformed speech using CycleGAN, and (e) reconstructed speech using CycleGAN.}
\label{fig:spectros}
\end{figure*}

\subsection{Subjective Evaluation}
\label{ssec:subj}

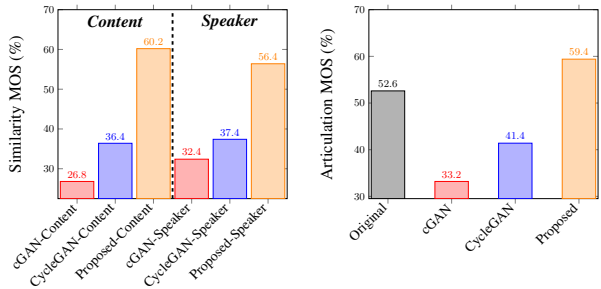
\begin{figure}[H]
\centering
\def\cscale{0.44}
\begin{subfigure}[t]{.23\textwidth}
\vskip 0pt
\begin{tikzpicture}[scale=\cscale]
  \begin{axis}[
    ybar=-1cm,
    bar width=1cm,
    xtick={cGAN-Content, CycleGAN-Content, Proposed-Content, cGAN-Speaker, CycleGAN-Speaker, Proposed-Speaker},
    symbolic x coords={cGAN-Content, CycleGAN-Content, Proposed-Content, cGAN-Speaker, CycleGAN-Speaker, Proposed-Speaker},
    x tick label style={rotate=45, anchor=east, align=left, font=\large},
    ylabel={\Large{Similarity MOS (\%)}},
    nodes near coords,
    nodes near coords align={vertical},
    ymax=70,
    ]
    \addplot[blue, fill=blue!30!white] coordinates {(CycleGAN-Content, 36.4)};
    \addplot[red, fill=red!30!white] coordinates {(cGAN-Content,  26.8)};
    \addplot[orange, fill=orange!30!white] coordinates {(Proposed-Content, 60.2)};
    \draw[ultra thick, dashed] ($(axis cs:Proposed-Content, \pgfkeysvalueof{/pgfplots/ymin})!.5!(axis cs:cGAN-Speaker, \pgfkeysvalueof{/pgfplots/ymin})$) -- ($(axis cs:Proposed-Content, \pgfkeysvalueof{/pgfplots/ymax})!.5!(axis cs:cGAN-Speaker, \pgfkeysvalueof{/pgfplots/ymax})$);
    \addplot[blue, fill=blue!30!white] coordinates {(CycleGAN-Speaker, 37.4)};
    \addplot[red, fill=red!30!white] coordinates {(cGAN-Speaker,  32.4)};
    \addplot[orange, fill=orange!30!white] coordinates {(Proposed-Speaker,  56.4)};
    \node at (axis cs:CycleGAN-Content,  67) {\Large{\textbf{\textit{Content}}}};
    \node at (axis cs:CycleGAN-Speaker,  67) {\Large{\textbf{\textit{Speaker}}}};
  \end{axis}
\end{tikzpicture}
\end{subfigure}
\begin{subfigure}[t]{.23\textwidth}
\vskip 0pt
\begin{tikzpicture}[scale=\cscale]
  \begin{axis}[
    ybar=-1cm,
    bar width=1cm,
    xtick={Original, cGAN, CycleGAN, Proposed},
    symbolic x coords={Original, cGAN, CycleGAN, Proposed},
    x tick label style={rotate=45, anchor=east, align=left, font=\large},
    ylabel={\Large{Articulation MOS (\%)}},
    nodes near coords,
    nodes near coords align={vertical},
    ymax=70
    ]
    \addplot[black, fill=black!30!white] coordinates {(Original,   52.6)};
    \addplot[blue, fill=blue!30!white] coordinates {(CycleGAN,  41.4)};
    \addplot[red, fill=red!30!white] coordinates {(cGAN,   33.2)};
    \addplot[orange, fill=orange!30!white] coordinates {(Proposed,   59.4)};
  \end{axis}
\end{tikzpicture}
\end{subfigure}
\caption{Mean opinion score (MOS) comparison}
\label{fig:mos}
\end{figure}

To determine whether these models improve intelligibility while preserving the content and speaker identity, we evaluate the MOS (mean opinion score) on CycleGAN, cGAN, and our model.
Figure~\ref{fig:mos} shows our MOS results.
Given the original utterance and the random shuffled utterances transformed by different models, five subjects are asked to evaluate the audio from three aspects: (1) how similar are the speaker characteristics before and after transformation (similarity-speaker); (2) how similar is the linguistic content before and after transformation (similarity-content); and (3) how clear is it compared to normal speech (articulation).
The percentiles of the MOS are normalized scores indicating the extent of which subjects consider samples to be similar or articulate respectively.

The similarity MOS indicates that our model does better than cGAN and CycleGAN in preserving both speaker characteristics and linguistic information.
CGAN performs the worst despite the additional use of ground truth information, because the amount of paired data is not sufficient to train the network.
In Section~\ref{ssec:stecyc} we further analyze why CycleGAN does not preserve the speaker characteristics and linguistic information.
The articulation MOS shows our improvement in intelligibility over impaired speech.
Audio samples for different approaches may be accessed at \url{https://b04901014.github.io/ISGAN/}.

\begin{figure}[H]
\centering
\begin{subfigure}{.22\textwidth}
  \centering
  \includegraphics[scale=0.205]{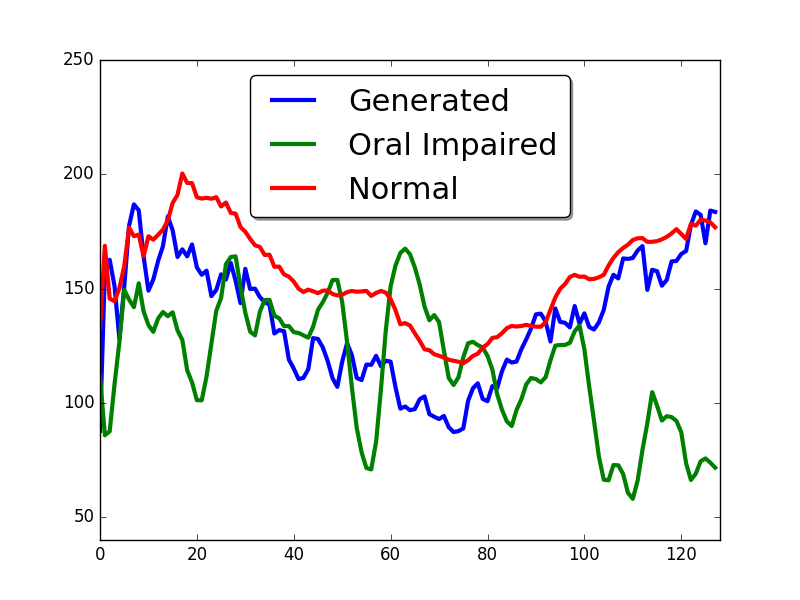}
  \caption{Global variance}
  \label{fig:GV}
\end{subfigure}
\begin{subfigure}{.22\textwidth}
  \centering
  \includegraphics[scale=0.205]{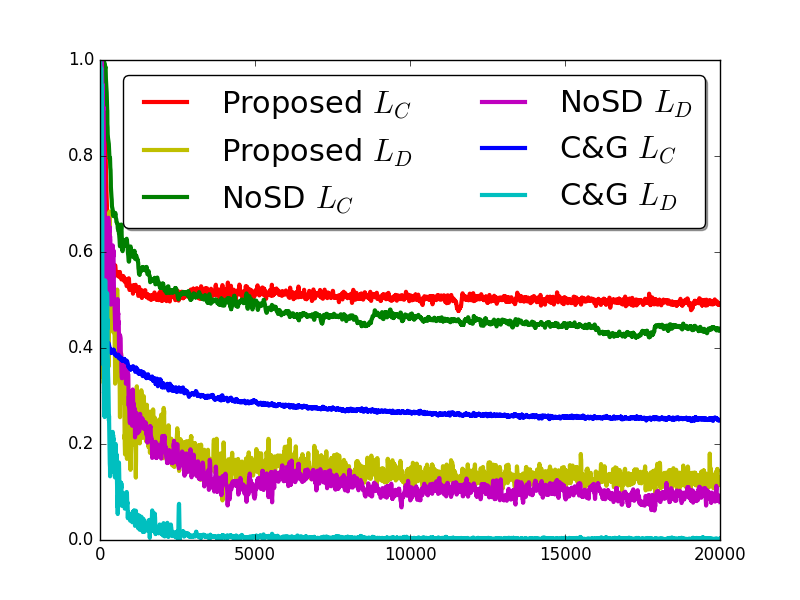}
  \caption{Loss curve}
  \label{fig:Losses}
\end{subfigure}
\caption{(a) GV before/after transformation by the proposed approach and normal utterances. 
(b) Loss curve of the models for the ablation study described in Section~\ref{ssec:abl}.}
\label{fig:expfigs}
\end{figure}

\subsection{Analysis of Proposed Approach}
\label{ssec:obj}

We first show the global variance (GV) for the proposed model.
As shown in Figure~\ref{fig:GV}, the GV of impaired speech  is quite different from that of normal speech.
The GV of the generated speech is similar to that of normal utterances.
Figures~\ref{fig:SpecOR} and \ref{fig:full} show an example of an impaired utterance and its transformed results using the proposed approach.
As shown in Figure~\ref{fig:SpecOR}, the orally impaired subject tends to have vague word boundaries, causing the continuous forms on the low frequency bands of the spectrogram.
Figure~\ref{fig:full} shows the ability to separate entangled word boundaries for the first and second word.
This can also be heard in the audio samples.
Nevertheless, we also see an obvious artifact around 2s in Figure~\ref{fig:full}.
This discontinuity is a consequence of our feeding the model of each time window independently, without any information from the previous window.
This may be solved in future work by feeding the previous window to the model as an augmented condition.

\subsection{Ablation Study}
\label{ssec:abl}

To show the contribution of each part of our model to training stability and audio quality, we conducted an ablation study.
We studied three different models: (i) the proposed model, (ii) the model without augmented input to the discriminator (NoSD)\footnote{In Figure~\ref{fig:Model}, the light blue arrows in the discriminator are removed.}, and (iii) the model in which the parameters of the generator $G$ and controller $C$ are updated to minimize both Equation~(\ref{eq:myc}) and~(\ref{eq:mylsgang}). That is, $G$ and $C$ are trained jointly without separate objectives (C\&G).
Figures~\ref{fig:Losses} and \ref{fig:SpecNSD} show the functionality of the augmented inputs of the discriminator.
The NoSD model gets both lower controller loss ($\mathcal{L}_C$) and discriminator loss ($\mathcal{L}_D$).
This indicates the controller is more capable of deceiving the generator, and as a consequence, the generator has less ability to generate plausible results to confuse the discriminator.
Thus the NoSD model yields a blurrier spectrogram in Fig.~\ref{fig:SpecNSD} than that for the proposed model in Fig.~\ref{fig:full}.
When updating $G$ and $C$ jointly (C\&G), Figure~\ref{fig:Losses} shows that the discriminator loss $\mathcal{L}_D$ quickly goes to zero, that is, the discriminator easily separates the real normal speech and generator output.
This indicates that the generator output can no longer be similar to the normal speech.

\subsection{Steganography of CycleGAN}
\label{ssec:stecyc}
As mentioned in \cite{CycSte}, CycleGAN learns to hide the information needed for reconstruction from the source domain into the target domain in an imperceptible manner.
We also see this phenomenon in Figures~\ref{fig:Cyctrans} and \ref{fig:Cycrec}.
The spectrograms before and after the CycleGAN transformation are quite different (Figure~\ref{fig:SpecOR} vs Figure~\ref{fig:Cyctrans}), whereas after transforming back to the source domain, the reconstructed audio is almost the same as the original input (Figure~\ref{fig:SpecOR} vs Figure~\ref{fig:Cycrec}).
This indicates that the cycle-consistency loss is not a good regularizer to enforce the model to have consistent input-output pairs.
Instead of using cycle-consistency loss, our method utilizes Equation~(\ref{eq:mygand1}) to maintain the consistency of content and speaker identity of the impaired and generated audio.
As shown in Figure~\ref{fig:mos}, the proposed approach is better. 

\section{Concluding Remarks}
Here we propose a novel unparallel VC model to improve the speech intelligibility of surgical patients who have had parts of their articulators removed.
In comparison with CycleGAN, which also needs only unparallel data, the proposed approach not only better improves articulation but also better preserves the linguistic content and speaker characteristics.

\bibliographystyle{IEEEtran}
\bibliography{refs,SLT18}

\end{document}